\documentclass[
prd,
superscriptaddress,
tightenlines,
showpacs,
showkeys,
nofootinbib,
aps,
amsfonts,
amssymb,
]{revtex4}

\usepackage{color}
\usepackage{amsmath}
\usepackage{amsfonts}
\usepackage{amssymb} 
\usepackage{graphicx}
\usepackage{epic}
\usepackage{eepic}
\usepackage{epsfig}
\usepackage{latexsym}
\usepackage{float}
\usepackage{shrthnds}
\usepackage[caption=false]{subfig}

\usepackage{geometry}
\begin{document}

\title{Probing Color Octet Electrons at the LHC}

\author{Tanumoy Mandal}
\email{tanumoy@imsc.res.in}
\affiliation{The Institute of Mathematical Sciences, Chennai, TN 600113, India}
\author{Subhadip Mitra}
\email{subhadip.mitra@th.u-psud.fr}
\affiliation{Laboratoire de Physique Th\'{e}orique, CNRS-UMR 8627, Universit\'{e} 
Paris-Sud 11, F-91405 Orsay Cedex, France}


\begin{abstract}

Models with quark and lepton compositeness predict the existence of colored 
partners of the Standard Model leptons. In this paper we study the LHC 
phenomenology of a charged colored lepton partner, namely the color octet electron, 
$e_8$ in an effective theory framework. We explore various mechanisms for resonant
production of $e_8$'s. With the pair production channel the 14 TeV LHC can probe
$e_8$'s with masses up to 2.5 TeV (2.8 TeV) with 100 fb$^{-1}$ (300 fb$^{-1}$) of
integrated luminosity. A common feature in all the resonant production channels 
is the presence of two high $p_T$ electrons and at least one high $p_T$ jet in 
the final state. Using this feature, we implement a search method where the signal 
is a combination of pair and single production events. This method has potential 
to increase the LHC reach significantly. 
Using the combined signal we estimate the LHC discovery potential for the $e_8$'s. Our
analysis shows that the LHC with 14 TeV center-of-mass energy and 100 fb$^{-1}$ 
(300 fb$^{-1}$) of integrated luminosity can probe  $e_8$'s with masses up to 3.4 TeV 
(4 TeV) for the compositeness scale of 5 TeV. 

\end{abstract}



\maketitle 

\section{Introduction}
\label{sec:intro}
 
All the experimental outcomes so far indicate that the Standard Model (SM) is 
the correct effective theory of elementary particles for energies below the TeV
scale. If the recently discovered Higgs-like boson at the Large Hadron Collider 
(LHC) at CERN is confirmed to be the SM Higgs, it will complete the experimental
verification of the particle spectrum of the SM \cite{HiggsAtlas,HiggsCMS}. 
However, despite its spectacular success with the experiments, there remain some
issues like the hierarchy problems, fermion family replication etc. that are not
properly addressed in the SM. Many theoretical attempts have been made to resolve
these issues. Beyond-the-SM (BSM) alternatives like supersymmetry, extra
dimensions and quark-lepton compositeness are some well-known examples. Many 
of the BSM theories predict the existence of new particles with masses near the TeV
scale. Two detectors of the LHC, namely ATLAS and CMS, are presently 
looking for the signatures of some of these new particles.

Of the various BSM scenarios, the quark-lepton composite models assume that 
the SM particles may not be fundamental and just as the proton has constituent
quarks, they are actually bound states of substructural constituents (preons)
\cite{Pati:1974yy}. These constituents are visible only beyond a certain energy
scale known as the compositeness scale. A typical consequence of quark-lepton 
compositeness is the appearance of colored particles with nonzero lepton numbers
(leptogluons, leptoquarks) and excited leptons etc. Some composite models naturally
predict the existence of leptogluons ($l_8$) \cite{Pati:1974yy,Terazawa:1976xx,
Neeman:1979wp,Harari:1979gi,Shupe:1979fv,Fritzsch:1981zh,D'Souza:1992tg} that 
are color octet fermions with nonzero lepton numbers. 
Several studies on the collider searches of leptoquarks and excited fermions can be
found in the literature \cite{Chatrchyan:2012dn,Chatrchyan:2012sv,ATLAS:2012af}
but there are only a few similar studies on $l_8$'s. Various signatures of 
color octet leptons at different colliders were investigated in some earlier 
papers \cite{Rizzo:1985dn,Rizzo:1985ud,Streng:1986my,Celikel:1998tm,
Hewett:1997ce,Celikel:1998dj}. Recently some important production processes 
of the $l_8$ have been analyzed for future colliders like the Large 
Hadron-electron Collider (LHeC), the International Linear Collider (ILC) and the
Compact Linear Collider (CLiC) \cite{Sahin:2010dd,Akay:2010sw}. We briefly 
review the limits on (charged) color octet leptons available in the literature. 
The lower mass limit of color octet charged leptons quoted in the latest 
Particle Data Book \cite{Beringer:2012pdg} is only 86 GeV. This limit is  
from the 23-years-old Tevatron data \cite{Abe:1989es} from the pair 
production channel. A mass limit of $M_{l_8} > \mathcal{O}(110)$ GeV from the
direct pair production via color interactions has been derived from $p\bar{p}$ 
collider data in \cite{Baur:1985ud}. Lower limits on the leptogluons masses were
derived by the JADE collaboration from the $t$-channel contribution to the total
hadronic cross section, $M_{l_8} \gtrsim (240~\textrm{GeV})^3/4\Lambda^2$ 
(where $\Lm$ is the compositeness scale) and from direct production via one 
photon exchange, $M_{l_8} \gtrsim 20$ GeV \cite{Bartel:1987zp}. In Ref.
\cite{Abt:1993nr}, the compositeness scale $\Lambda \lesssim 1.8$ TeV was 
excluded at 95$\%$ confidence level (CL) for $M_{l_8} \simeq 100$ GeV and 
$\Lambda \lesssim 200$ GeV for $M_{l_8} \simeq 200$ GeV. It is also mentioned 
in \cite{Hewett:1997ce} that the D0 cross section bounds on $eejj$ events 
exclude leptogluons with masses up to 200 GeV and could naively place the constraint
$M_{l_8} \gtrsim 325$ GeV.

In this paper we study the LHC discovery potential for a generic color octet 
partner of a charged lepton, namely the color octet electron, $e_8$. Although, 
in this paper, we consider only color octet electrons, our results are applicable 
for the color octet partner of the muon, {\it i.e.}, $\m_8$ also.  
The paper is organized as follows: in Sec. \ref{sec:lag} we display the
interaction Lagrangian and decay width of $e_8$. In Sec. \ref{sec:LHCsig} we
discuss various $e_8$ production processes at the LHC. In Sec. \ref{sec:LHCdis} 
we discuss the LHC reach for $e_8$'s. Finally, in Sec. \ref{sec:conclu} we
summarize and offer our conclusions.


\section{The Lagrangian}
\label{sec:lag}

Assuming lepton flavor conservation we consider a general Lagrangian for the 
color octet electrons including terms allowed by the gauge symmetries of the SM,
\ba
\lag = \bar{e}_8^a i\g^\m\left(\pr_\m\dl^{ac} + g_s f^{abc}G^b_\m \right) e_8^c 
- M_{e_8}\bar{e}_8^a e_8^a + \lag_{int}.\label{eq:lag}
\ea
For simplicity, we have ignored the terms with electroweak couplings. The
interaction part $(\lag_{int})$ contains all the higher dimensional operators. In this paper 
we consider only the following dimension five terms that contain the interaction
between ordinary electrons and color octet ones \cite{Beringer:2012pdg},
\footnote{There are actually more dimension five operators allowed by the gauge
symmetries and lepton number conservation like,
\bas
\frac{\mc C_8}{\Lm} i f^{abc} \bar{e}_8^a G_{\m\n}^{b} \s^{\m\n} e_8^c +  
\frac{\mc C_1}{\Lm} \bar{e}_8^a B_{\m\n} \s^{\m\n} e_8^a.
\eas
However, these terms lead to momentum dependent $e_8e_8V$ vertices (form factors). 
Moreover, the octet term can lead to an $e_8 e_8 g g$ vertex which can affect 
the production cross section. We assume the unknown coefficients associated 
with these terms are negligible.}
\ba
\lag_{int} = \frac{g_s}{2\Lm} G^a_{\m\n} \left[\bar{e}_8^a \s^{\m\n}\left(\et_L e_{L} + \et_R e_{R}\right)\right] + H.c.\,.\label{eq:intlag}
\ea
Here $G^a_{\m\n}$ is the gluon field strength tensor, $\Lm$ is the scale below which  
this effective theory is valid and $\et_{L/R}$ are the chirality factors. 
Since, electron chirality conservation implies $\et_L\et_R = 0$, we set 
$\et_L = 1$ and $\et_R=0$ in our analysis.

From the interaction Lagrangian given in Eq. \ref{eq:intlag} we see that an $e_8$ can decay to a gluon and an electron (two-body decay mode), 
{\it i.e.}, $e_8 \to e g$  or two gluons and an electron (three-body decay mode), {\it i.e.}, $e_8 \to e gg$ (via the $e_8egg$ vertex). However, compared to the $e_8eg$ decay rate, the three-body decay rate is suppressed by an extra power of $\alpha_s$ and phase space suppression. If one includes dimension six or higher dimensional terms in the Lagrangian then, in general, $e_8$ can have other many-body decay modes like,  {\it e.g.}, $e_8\to eqq$ via a four fermion $e_8 eqq$ vertex. However, these many-body decays will be much more suppressed than the two-body decay. Hence, in this paper, we focus only on the dominant two-body decay mode. With $\et_L = 1$ and $\et_R=0$, the decay width of 
$e_8$ can be written as,
\ba
\G_{e_8} = \frac{\al_s(M_{e_8})M_{e_8}^3}{4\Lm^2}.\label{eq:width} 
\ea


\section{Production at the LHC}
\label{sec:LHCsig}		

In this section we discuss various production mechanisms of $e_8$'s at the LHC
and present the production cross sections for different channels. To obtain 
the cross sections, we have first implemented the Lagrangian of Eq. \ref{eq:lag} 
in FeynRules version 1.6.0 \cite{Christensen:2008py} to generate Universal
FeynRules Output (UFO) \cite{Degrande:2011ua} format model files suitable for
MadGraph5 \cite{Alwall:2011uj} that we have used to estimate the cross sections. 
We have used CTEQ6L1 Parton Distribution Functions (PDFs) \cite{Pumplin:2002vw} 
for all our numerical computations.

At a hadron collider like the LHC, resonant productions of $e_8$'s can occur 
via $gg$, $gq$ and $qq$ initiated processes where $q$ can be either a light quark
or a bottom quark. The gluon PDF dominates at the low $x$ region whereas the quark 
PDFs take over at the high $x$ region. Thus, depending on $M_{e_8}$, all of the $gg$,
$gq$ and $qq$ initiated processes can contribute significantly to the production 
of $e_8$'s at the LHC.

For the resonant production $e_8$'s at colliders, two separate channels are
generally considered in the literature -- one is the pair production
\cite{Hewett:1997ce,Celikel:1998dj} and the other is the single production of 
$e_8$ \cite{Rizzo:1985dn,Rizzo:1985ud,Streng:1986my,Celikel:1998tm,
Sahin:2010dd}. In general, pair production of a colored particle is considered 
mostly model independent. This is because the universal strong coupling constant
$g_S$ controls the dominant pair production processes unlike the single production
processes where the cross section depends more on various model parameters like
couplings and scales etc. However, as we shall see, for $e_8$'s, the $t$-channel
electron exchange diagrams can contribute significantly to the pair production
making it more model dependent.
 
 \subsection{Pair Production (${\bf gg,qq \to e_{8}e_{8}}$)}
\label{subsec:e8e8}

\begin{figure}[!h]
\centering
\subfloat{
\begin{tabular}{ccccccc}
\resizebox{35mm}{!}{\includegraphics{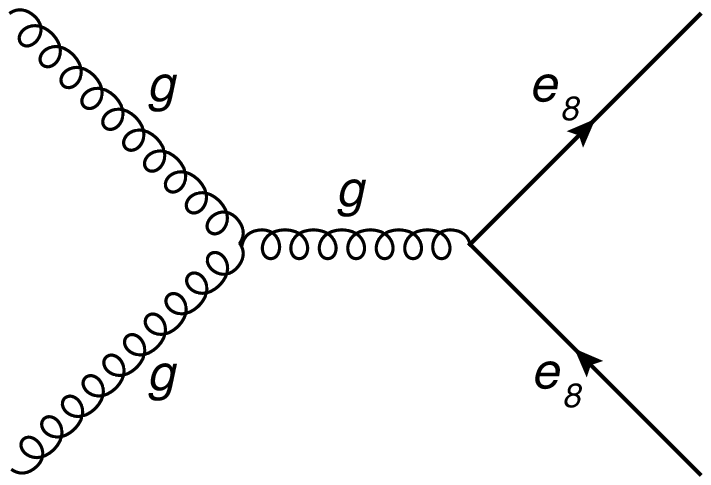}} &&
\resizebox{35mm}{!}{\includegraphics{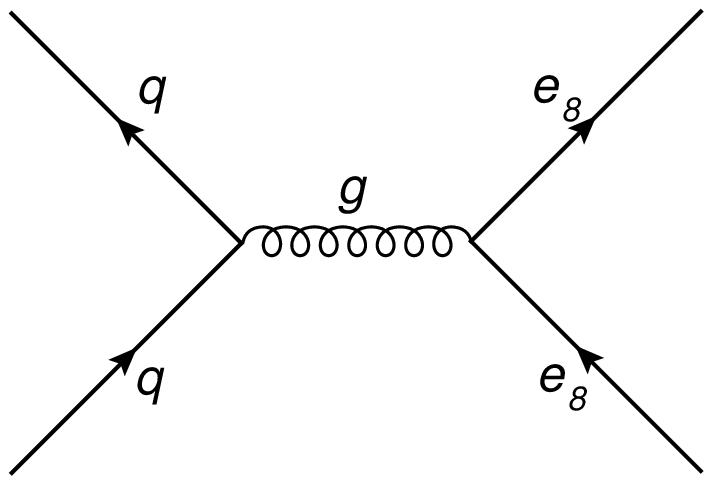}} &&
\resizebox{35mm}{!}{\includegraphics{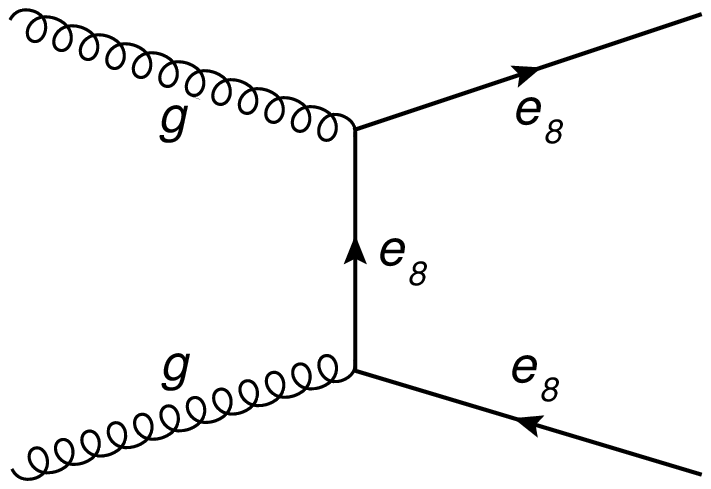}} &&
\resizebox{35mm}{!}{\includegraphics{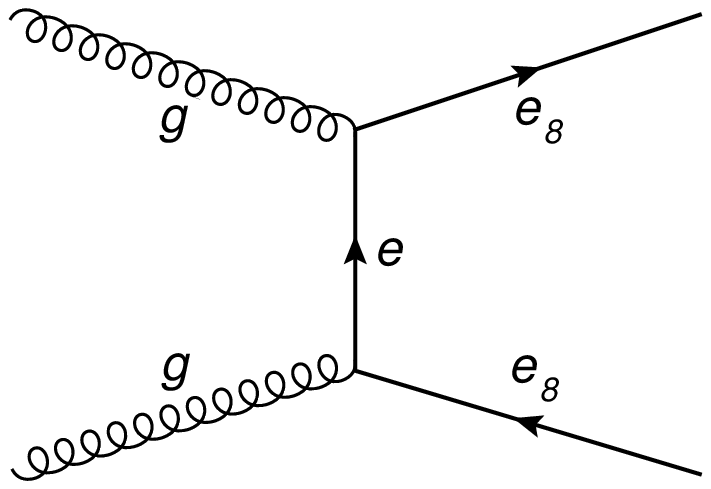}}\\
(a)&&(b)&&(c)&&(d)
\end{tabular}}\\
\caption{\label{fig:e8e8FD}Parton level Feynman diagrams for $pp\to e_8e_8$
processes at the LHC.}
\end{figure}

\begin{figure}[!h]
\centering
\subfloat{
\begin{tabular}{c}
\resizebox{100mm}{!}{\includegraphics{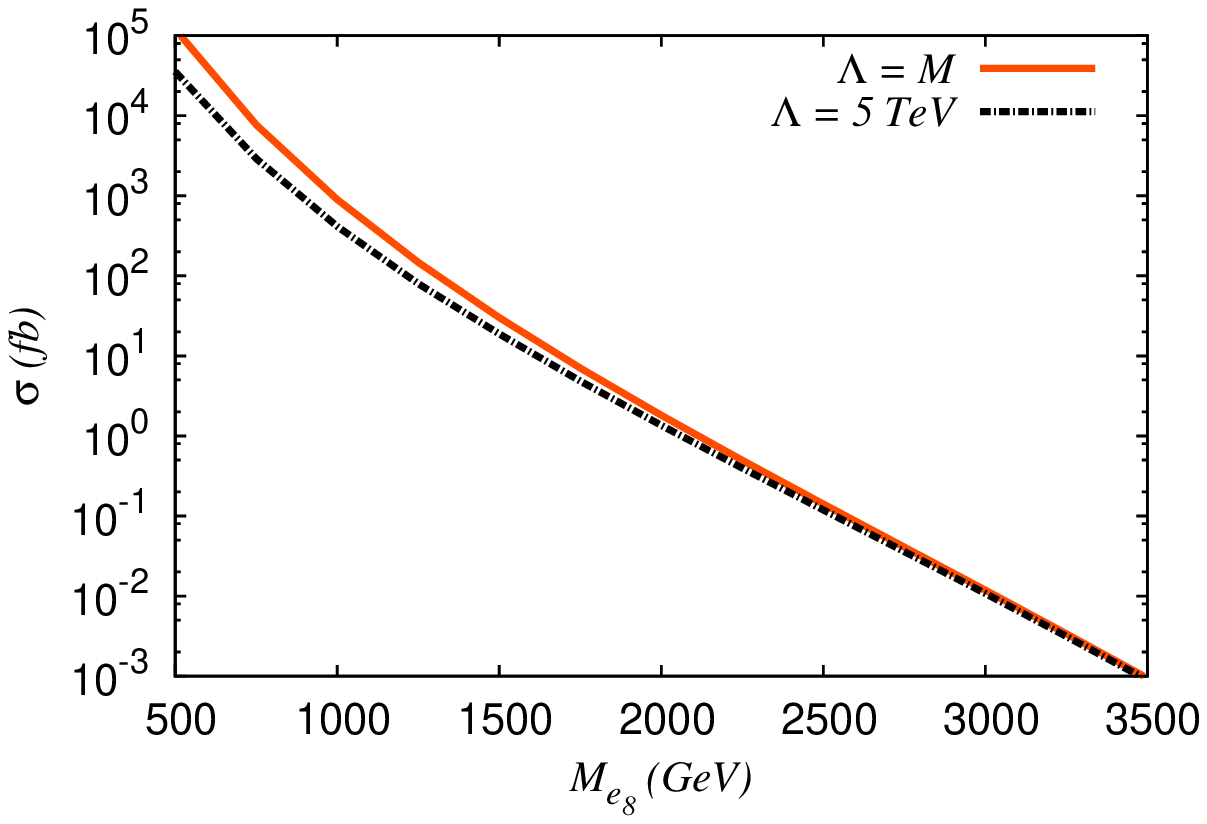}}\\
\end{tabular}}
\caption{\label{fig:pairCS} The cross sections for $pp\to e_8e_8$ as a function of 
$M_{e_8}$ for $\Lm = M_{e_8}$ and $\Lm =5$ TeV at the 14 TeV LHC.}
\end{figure}

\begin{figure}[!h]
\centering
\subfloat{
\begin{tabular}{c}
\resizebox{100mm}{!}{\includegraphics{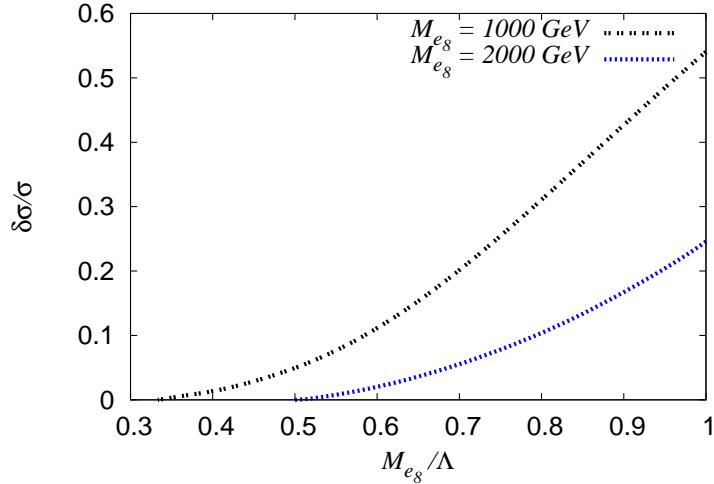}}\\
\end{tabular}}
\caption{\label{fig:model_dep} Dependence of $\dl\s/\s$ (defined in Eq. \ref{eq:model_dep}) on $M_{e_8}/\Lm$ for $M_{e_8}=1$ TeV and 2 TeV
at the 14 TeV LHC.}
\end{figure}

At the LHC, pair production of $e_8$'s is $gg$ or $qq$ initiated -- see Fig.
\ref{fig:e8e8FD} where we have shown the parton level Feynman diagrams for this
channel. Of these, only the electron exchange diagram, shown in 
Fig \ref{fig:e8e8FD}d, contains the 
$\Lm$ dependent $gee_8$ vertex. In Fig. \ref{fig:pairCS} we show the 
$pp\to e_8e_8$ cross section as a function of $M_{e_8}$ for two different 
choices of $\Lm$, $\Lm=M_{e_8}$ and $\Lm = 5$ TeV, at the 14 TeV LHC. 
In Fig. \ref{fig:model_dep} we have plotted $\dl\s$ as a function of $\Lm$ 
to show the dependence of the pair production cross section on $\Lm$ for 
$M_{e_8} = 1$ and 2 TeV, where $\dl\s$ is a measure of the contribution of
the electron exchange diagram and is defined as,
\ba
\dl\s (\Lm) = \s (\Lm) - \s (\Lm\rar\infty).\label{eq:model_dep}
\ea

As $\Lm$ increases the contribution coming from the electron exchange diagrams
decreases and for $\Lm \gg M_{e_8}$ becomes negligible. So the pair production 
is model independent only for very large $\Lm$. 

After being produced as pairs at the LHC, each $e_8$ decays into an electron 
(or a positron) and a gluon at the parton level, {\it i.e.}, 
\begin{equation*}
gg/qq\to e_8e_8 \to eejj.
\end{equation*} 
For large $M_{e_8}$, these two jets and the lepton pair will have high $p_{T}$.
This feature can be used to isolate the $e_8$ pair production events from the 
SM backgrounds at the LHC.

\subsection{Two-body Single Production (${\bf gg,qq \to e_{8}e}$)}
\label{subsec:e8e}

\begin{figure}[!h]
\centering
\subfloat{
\begin{tabular}{ccccccc}
\resizebox{35mm}{!}{\includegraphics{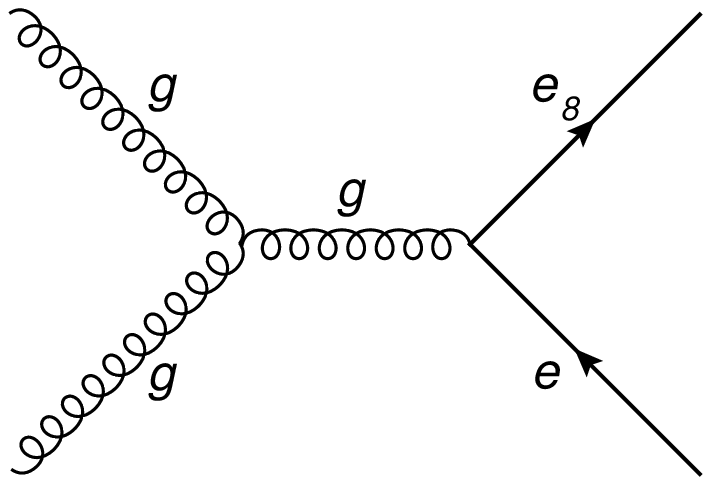}} &&
\resizebox{35mm}{!}{\includegraphics{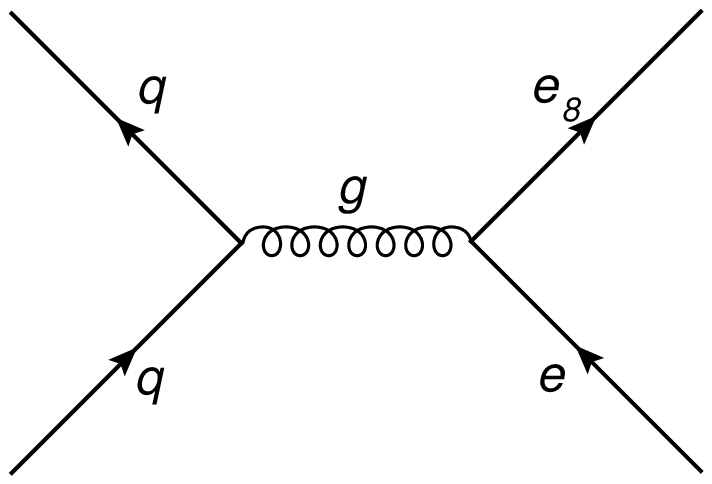}} &&
\resizebox{35mm}{!}{\includegraphics{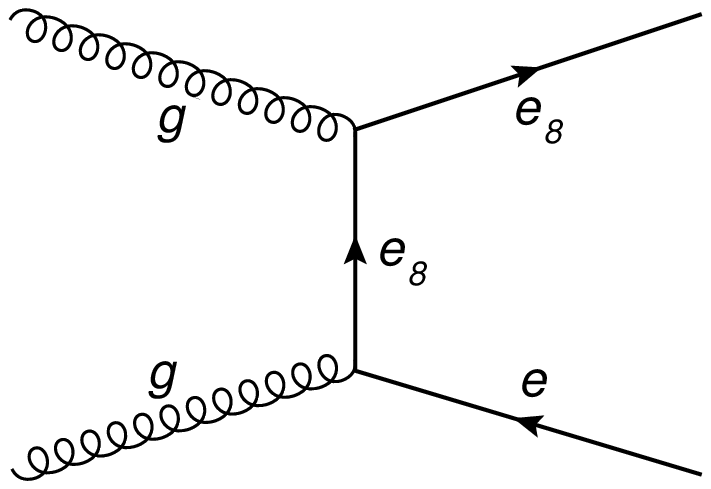}} &&
\resizebox{35mm}{!}{\includegraphics{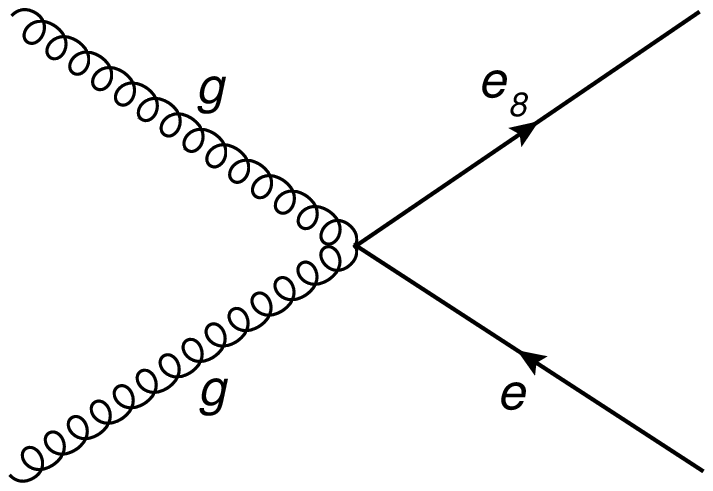}}\\
(a)&&(b)&&(c)&&(d)
\end{tabular}}\\
\caption{\label{fig:e8eFD}Parton level Feynman diagrams for $pp\to e_8e$
processes at the LHC.}
\end{figure}

\begin{figure}[!h]
\centering
\subfloat{
\begin{tabular}{c}
\resizebox{100mm}{!}{\includegraphics{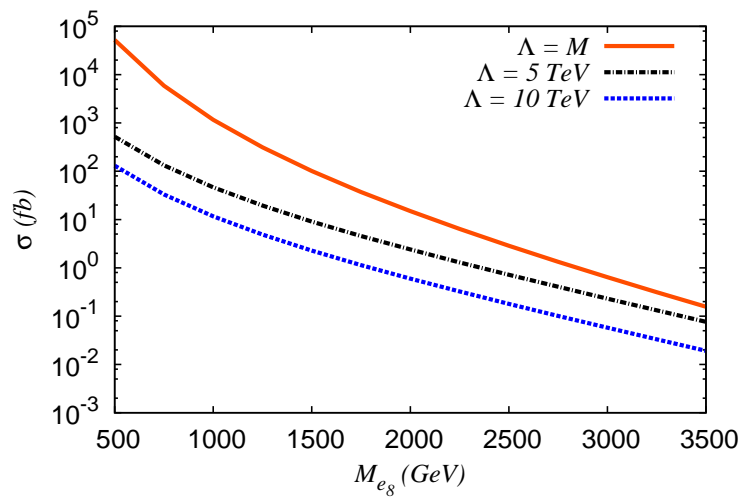}} \\
\end{tabular}}
\caption{\label{fig:single2bCS} The cross sections for $pp\to e_8e$ as a function
of $M_{e_8}$ for $\Lm = M_{e_8}$, 5 TeV and 10 TeV at the 14 TeV LHC.}
\end{figure}

The two-body single production channel where an $e_8$ is produced in association
with an electron can have either $gg$ or $qq$ initial states as shown 
in Fig. \ref{fig:e8eFD}. This channel is model dependent as each Feynman diagram
for the $pp\to e_8e$ process contains a $\Lm$ dependent vertex. In Fig.
\ref{fig:single2bCS} we show the $pp\to e_8e$ cross sections as a function of
$M_{e_8}$ with $\Lambda = M_{e_8}$ and 5 TeV and 10 TeV at the 14 TeV LHC.

As the $e_8$ decays, this process gives rise to an $eej$ final state at the 
parton level. The $e$ and the $j$ produced from the decay of the $e_8$, have 
high $p_T$. The other $e$ also possesses very high $p_T$ as it balances against 
the massive $e_8$. 

\subsection{Three-body Single Production (${\bf gg,gq,qq \to e_{8}ej}$)}
\label{subsec:e8ej}

\begin{figure}[!h]
\centering
\subfloat{
\begin{tabular}{ccccccc}
\resizebox{35mm}{!}{\includegraphics{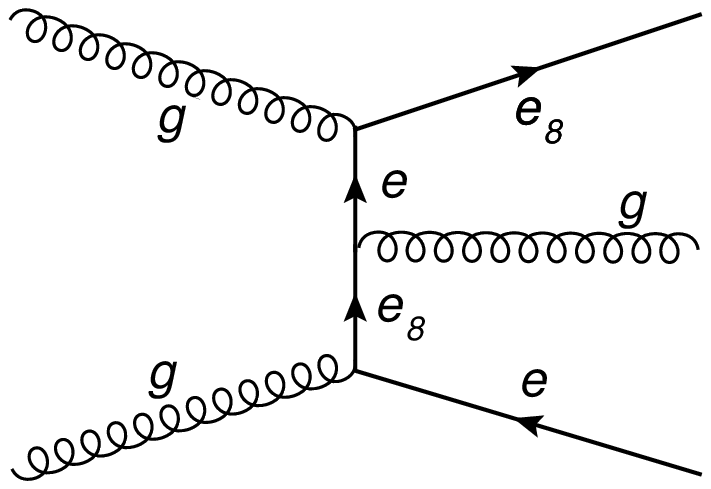}} &&
\resizebox{35mm}{!}{\includegraphics{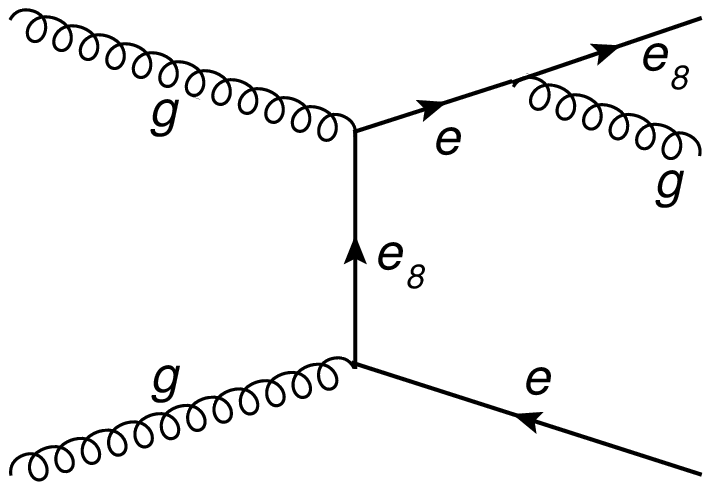}} &&
\resizebox{35mm}{!}{\includegraphics{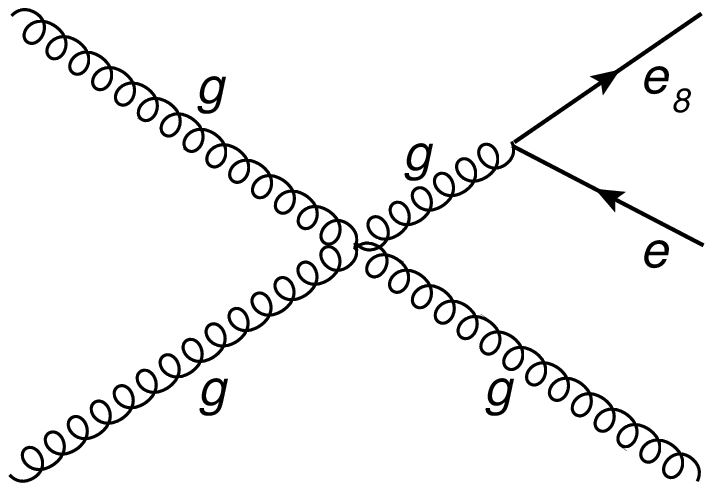}} &&
\resizebox{35mm}{!}{\includegraphics{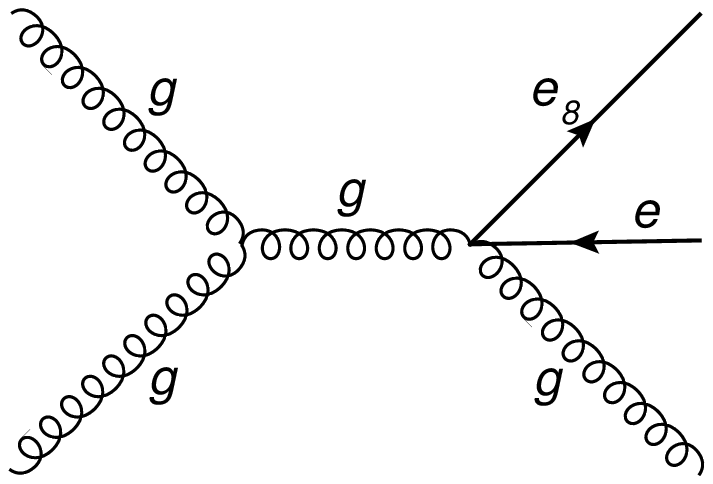}} \\
(a)&&(b)&&(c)&&(d)\\
\vspace{0.4cm}\\
\resizebox{35mm}{!}{\includegraphics{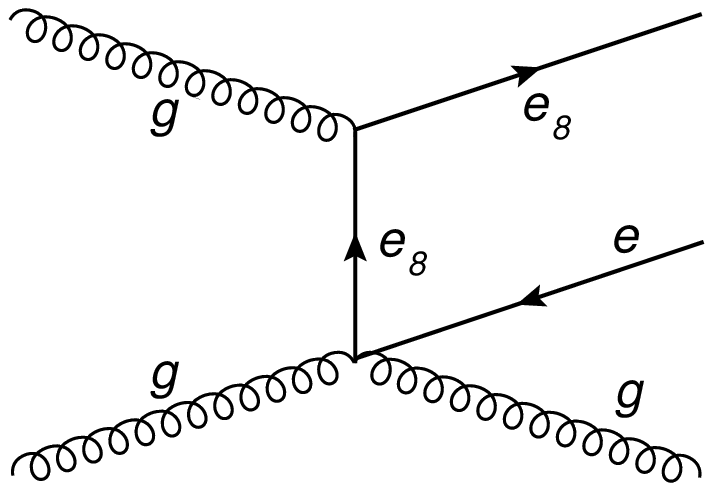}} &&
\resizebox{35mm}{!}{\includegraphics{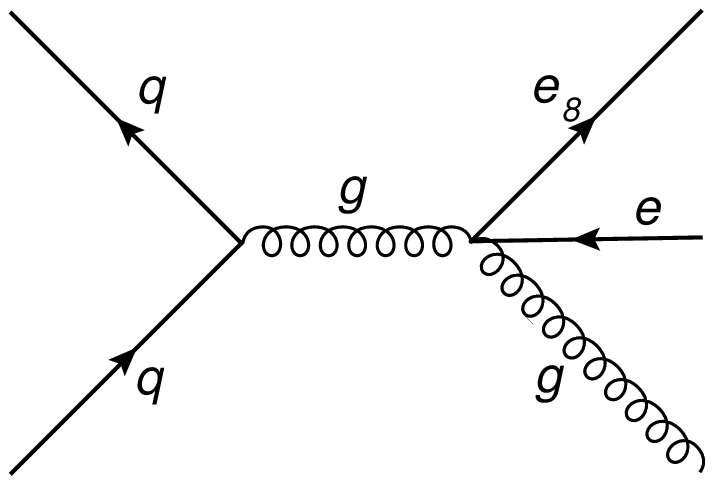}} &&
\resizebox{35mm}{!}{\includegraphics{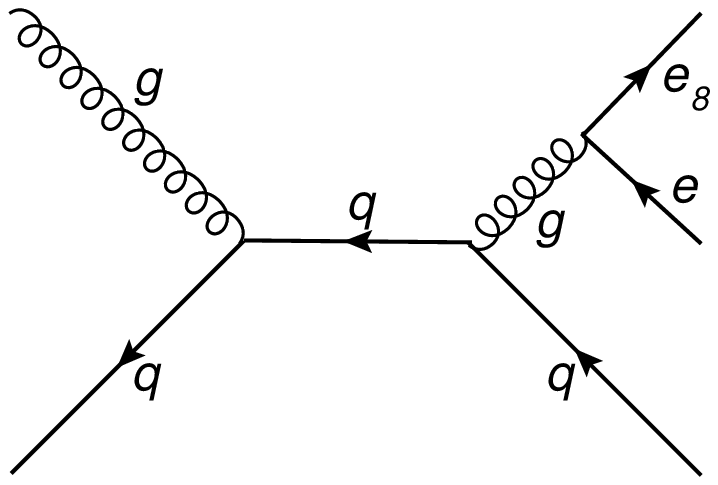}} &&
\resizebox{35mm}{!}{\includegraphics{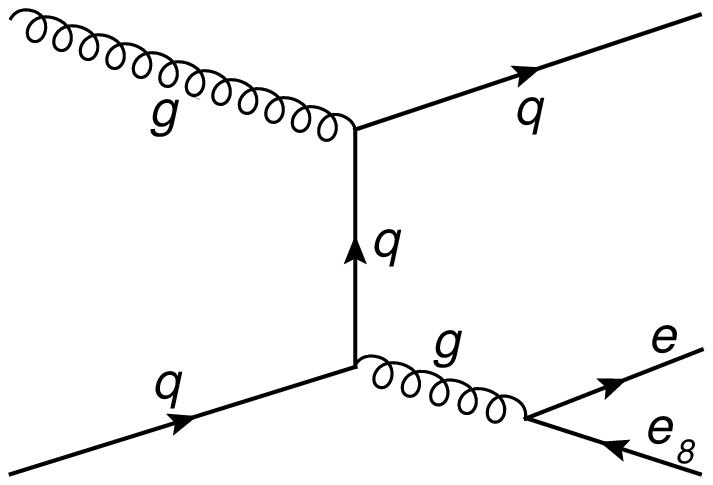}} \\
(e)&&(f)&&(g)&&(h)\\
\vspace{0.4cm}\\
\resizebox{35mm}{!}{\includegraphics{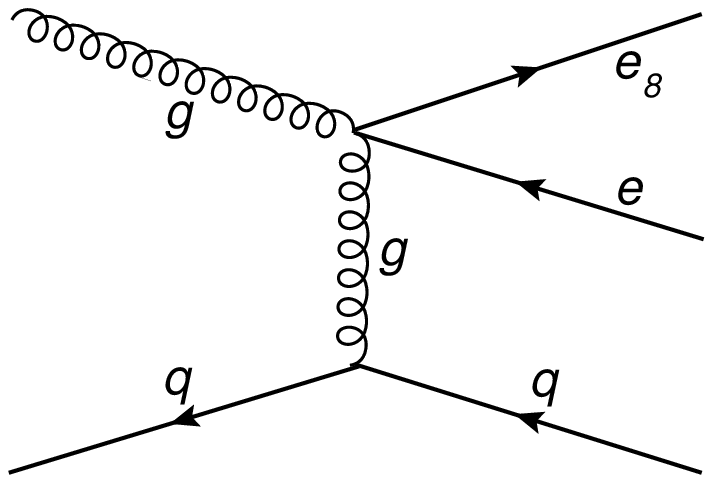}} &&
\resizebox{35mm}{!}{\includegraphics{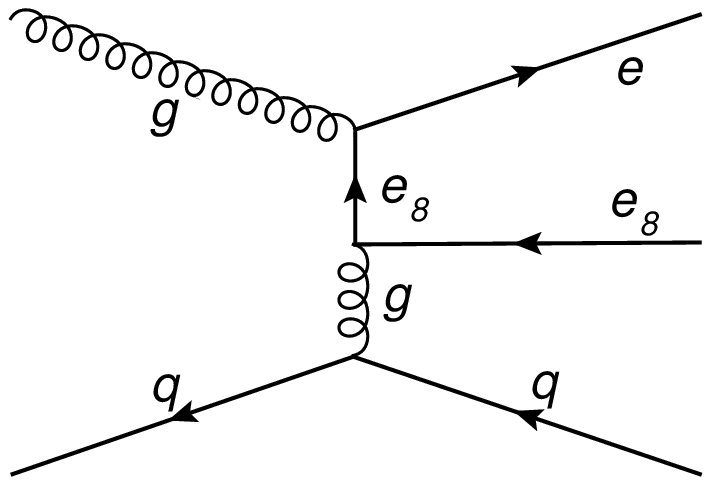}} &&
\resizebox{35mm}{!}{\includegraphics{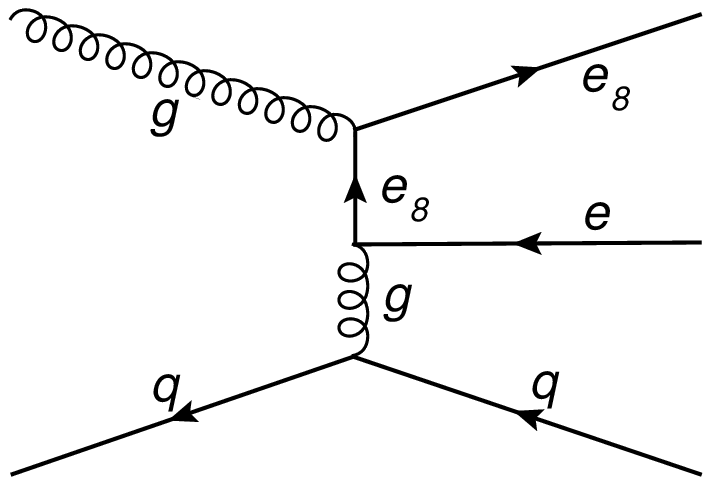}} &&
\resizebox{35mm}{!}{\includegraphics{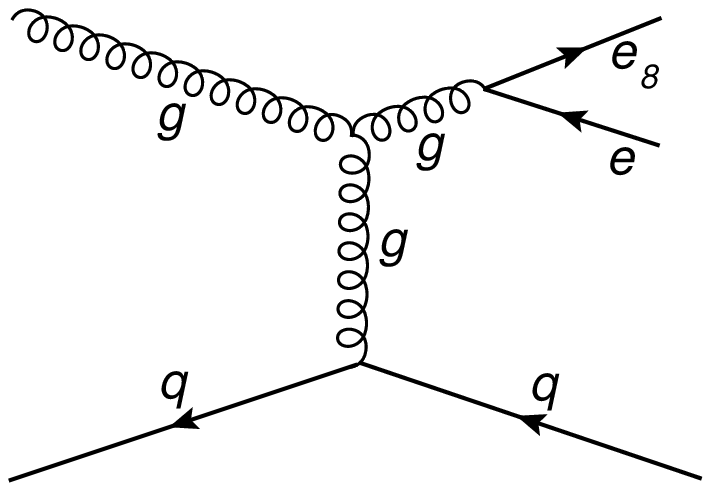}}\\
(i)&&(j)&&(k)&&(l)
\end{tabular}}\\
\caption{\label{fig:e8ejFD}Parton level Feynman diagrams for $pp\to e_8ej$
processes of the third type at the LHC.}
\end{figure}

Apart from the pair and the two-body single productions, we also consider single
production of an $e_8$ in association with an electron and a jet. The 
$pp \to e_8ej$ process includes three different types of diagrams as follows:

\begin{enumerate}

\item 
The diagrams where the $ej$ pair is coming from another $e_8$. 
Though there are three particles in 
the final state, this type of diagram effectively corresponds to two-body pair
production process.

\item
The two-body single production ($pp \to e_8e$) process with a jet radiated from the
initial state (ISR) or final state (FSR) or intermediate virtual particles can 
lead to an $e_8ej$ final state.
  
\item
A new set of diagrams that are different from the two types of diagrams mentioned
above. These new channels can proceed through $gg$, $qq$ and $gq$ initial states 
as shown in Fig. \ref{fig:e8ejFD}.

\end{enumerate}

This new set of diagrams has not been considered so far in the literature. 
It is difficult to compute the total contribution of these diagrams 
in a straight forward manner with a leading order parton level matrix element
calculation because of the presence of soft radiation jet emission diagrams. 
In order to get an estimation of the contribution of these new diagrams without
getting into the complicacy of evaluating the soft jet emission diagrams,
here, in this section, we present the cross section only for the $gq$ initiated
processes, {\it i.e.} $gq \to e_8ej$ since the first and the second types of 
diagrams can not be initiated by $gq$ state. In Fig. \ref{fig:compare} we
show the cross section of the $gq \to e_8ej$ process along with the $pp \to e_8e_8$
and the $pp \to e_8e$ processes. We find that the cross sections even for the 
$gq$ initiated subset can be comparable to the $pp \to e_8e_8/e_8e$ processes for 
large $M_{e_8}$ despite the facts that these new diagrams have three-body final
states and are suppressed by one extra power of the coupling (either $g_S$ or
$g_S/\Lm$) compared to the two-body single and pair production processes. However,
since there is one less $e_8$ compared to the pair production process, depending 
on the coupling the three-body phase space of the single production can be 
comparable or even larger to the two-body phase space of the pair production 
for large $M_{e_8}$.

After the $e_8$ decay, the three-body single production process is characterized 
by an $eejj$ final state like the pair production. However, unlike the pair
production, here one of the jets can have a low transverse momentum. 

\subsection{Indirect Production (${\bf gg \to ee}$)}
\label{subsec:ee}

\begin{figure}[!h]
\centering
\subfloat{
\begin{tabular}{ccc}
\resizebox{75mm}{!}{\includegraphics{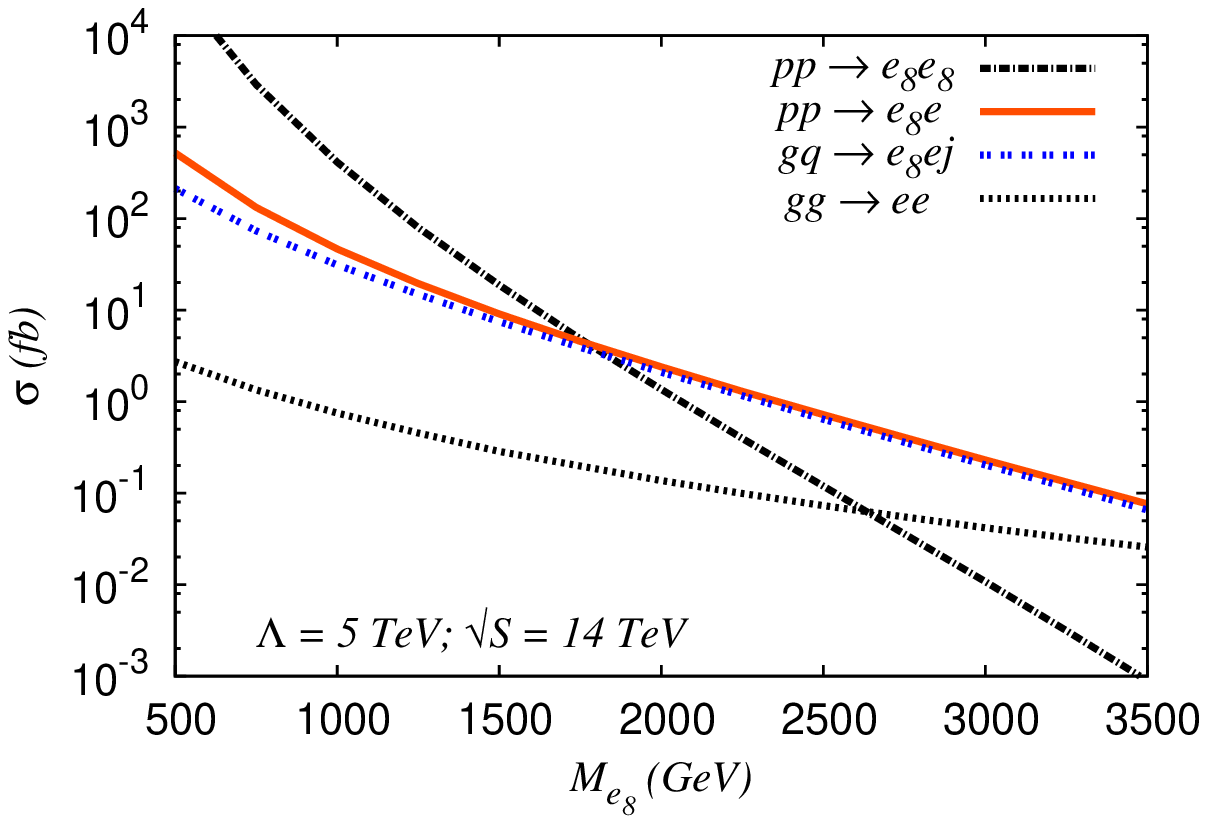}} &&
\resizebox{75mm}{!}{\includegraphics{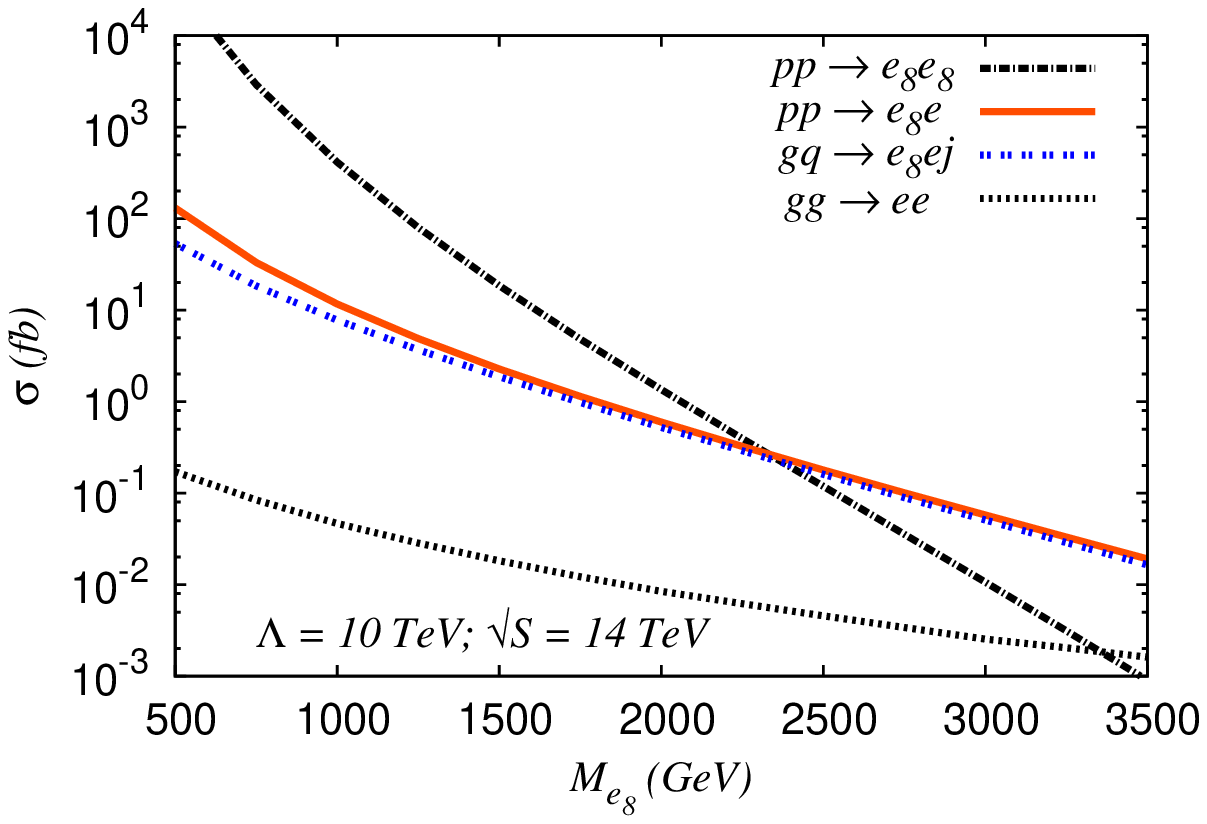}} \\
\footnotesize{\hspace{1.1cm}(a)}&&\footnotesize{\hspace{1.2cm}(b)}
\end{tabular}}
\caption{\label{fig:compare}Cross sections for $pp\to e_8e_8$, $pp\to e_8e$,
$gq\to e_8ej$ and $gg\xrightarrow{e_8} ee$ processes for $\Lm =5$ TeV and 10 TeV 
at the 14 TeV LHC. The $\sigma(gq \to e_8ej)$ is computed with the following
kinematical cuts: $p_T(j)>25$ GeV and $|y(j)|<2.5$.}
\end{figure}

\begin{figure}[!h]
\centering
\subfloat{
\begin{tabular}{c}
\resizebox{40mm}{!}{\includegraphics{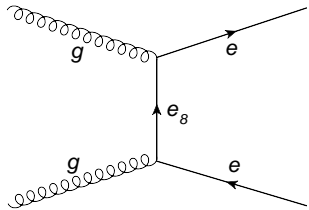}}
\end{tabular}}\\
\caption{\label{fig:gg2eeFD}Parton level Feynman diagram for indirect production of $e_8$'s at the LHC.}
\end{figure}

So far we have considered only resonant production of $e_8$'s. However, a 
t-channel exchange of the $e_8$ can convert a gluon pair to an electron-positron
pair at the LHC (Fig. \ref{fig:gg2eeFD}). Similar indirect productions in the
context of future linear colliders such as the ILC and CLiC have been 
analyzed in \cite{Akay:2010sw}. Indirect production is less significant because 
the amplitude is proportional to $1/\Lm^2$. Moreover, at the LHC this is also 
color suppressed because of the color singlet nature of the final states.
In Fig. \ref{fig:compare} we also show the cross section of the indirect 
production process at the LHC.


\section{LHC Discovery Potential}
\label{sec:LHCdis}

From Fig. \ref{fig:compare} we see that for small $M_{e_8}$, the pair production
cross section is larger than the other channels. As $M_{e_8}$ increases, it
decreases rapidly due to phase-space suppression and the single production 
channels (both the two-body and the three-body) take over the pair production (the
crossover point depends on $\Lm$).  Hence, if $\Lm$ is not too high,  
the single production channels will have better reach than the pair production
channel and so, to estimate the LHC discovery reach, we consider both the pair 
and the single production channels. However, while estimating for the single
production channels we have  to remember that because of the radiation jets, 
it will be difficult to separate the two-body and the three-body single
productions at the LHC. So, in this paper, we consider a selection criterion 
that combines events from all the production processes at the LHC.

\subsection{Combined Signal}
To design the selection criterion mentioned above we first note some of the
characteristics of the final states of the resonant production processes
\footnote{We focus on the resonant productions because as we saw the indirect
production is less significant at the LHC.}, 

\begin{enumerate}

\item 
Process $pp \to e_8e_8 \to (eg)(eg)$ has two high $p_T$ electrons and two
high $p_T$ jets in the final state.

\item
Process $pp \to e_8e \to (eg)e$ has two high $p_T$ electrons and one
high $p_T$ jet in the final state.

\item
Process $pp \to e_8ej \to (eg)ej$ has two high $p_T$ electrons and at least one
high $p_T$ jet in the final state.

\end{enumerate}

All these processes have one common feature that they have two high $p_T$ 
electrons and a high $p_T$ jet in the final state. Hence if we demand 
that the signal events should have two high $p_T$ electrons and at least one
high $p_T$ jet, we can capture events from all the above mentioned production
processes. To estimate the number of signal events that pass the above selection
criterion we combine the events from all the production channels mentioned in 
the previous section. However, as already pointed out, it is difficult to estimate
the number of signal events  with only a matrix element (ME) level Monte Carlo
computation due to the presence of soft
radiation jets. Hence we use the MadGraph ME generator to compute the hard 
part of the amplitude and Pythia6 (via the MadGraph5-Pythia6 interface) for 
parton showering. We also match the matrix element partons with the parton
showers to estimate the inclusive signal without double counting 
(see the Appendix \ref{matching} for more details on the matched signal).

\subsection{SM Backgrounds}

With the selection criterion mentioned above, the SM backgrounds are characterized
by the presence of two opposite sign electrons and at least one jet in the final
state. At the LHC, the main source of $e^+e^-$ pairs (with high $p_T$) is the $Z$
decay \footnote{Here we don't include $e^+e^-$ pairs that come from $\g^*$. 
However, as we shall demand very high $p_T$ for both the electrons,
this background becomes negligible and won't affect our results too much.}.
Hence we compute the inclusive $Z$ production as the main background. Here, too, 
we compute this by matching the matrix element partons of $Z+ n$ jets 
($n = 0,1,2,3$) processes\footnote{Here $pp\to Zjj$ includes the processes 
where the jets are coming from a $W$ or a $Z$.} with the parton showers using the
shower-$k_T$ scheme \cite{Alwall:2008qv}. For the background, we also consider
some potentially significant processes to produce $e^+e^-$ pairs,
\ba
&& pp\to tt \to (bW)(bW) \to ( b e \n_e) (b e \n_e),\nn\\
&& pp \to tW \to bWW \to (b e \n_e) (e \n_e),\nn\\
&& pp \to WW \to (e \n_e) (e \n_e)\nn.
\ea

Note that all these processes have missing energy because of the $\n_e$'s in 
the final state. In Table \ref{tab:SMBG} we show the relative contributions of
these backgrounds generated with some basic kinematical cuts (to be described
shortly) on the final states . As mentioned, we see in Table \ref{tab:SMBG} 
that the inclusive $Z$ contribution overwhelms the other background processes.

\begin{figure}[!h]
\centering
\subfloat{
\begin{tabular}{ccc}
\resizebox{65mm}{!}{\includegraphics{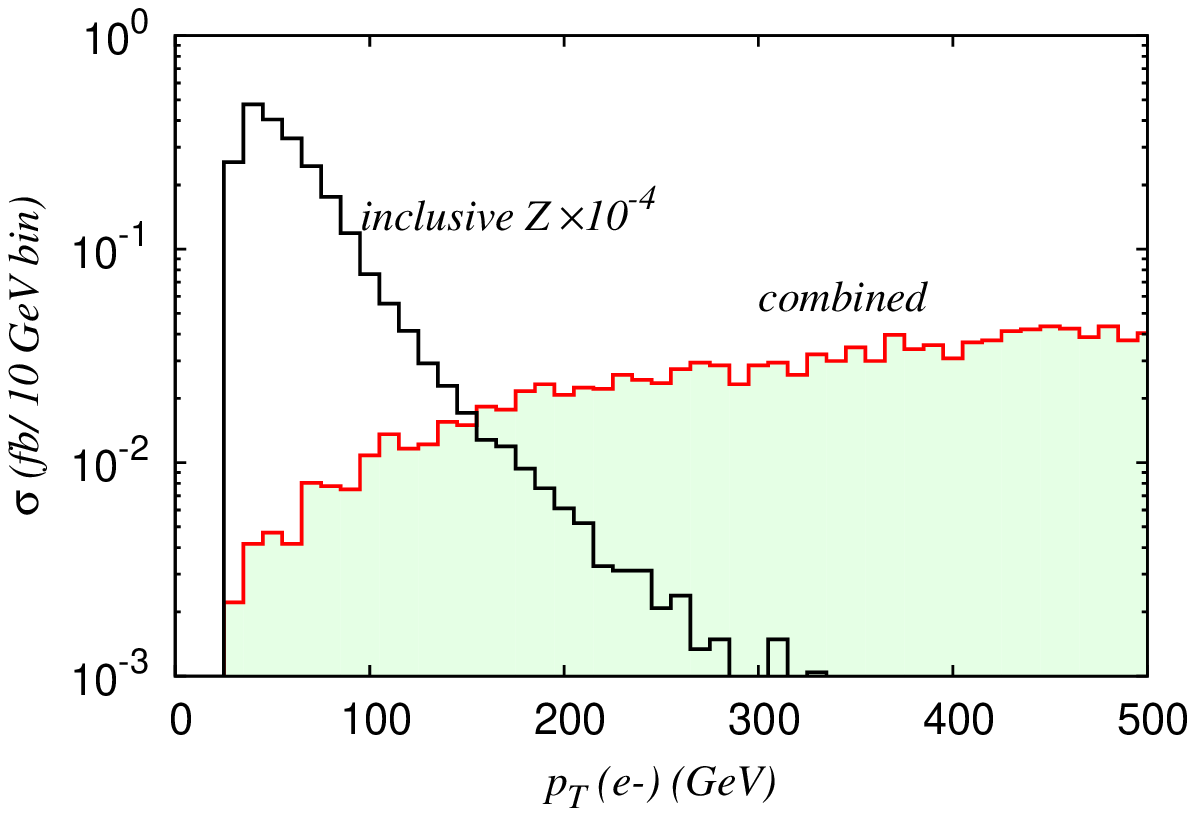}} &&
\resizebox{65mm}{!}{\includegraphics{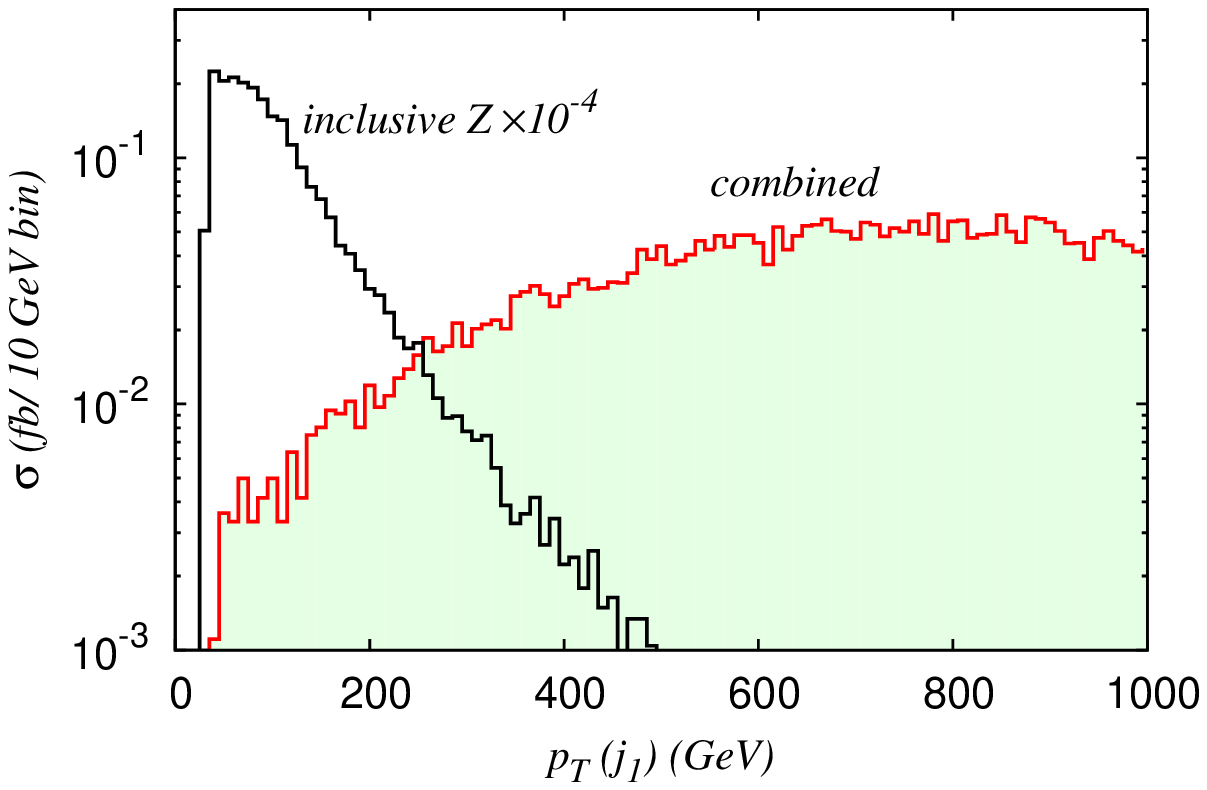}} \\
\footnotesize{\hspace{1.1cm}(a)}&&\footnotesize{\hspace{1.2cm}(b)}\\
\resizebox{65mm}{!}{\includegraphics{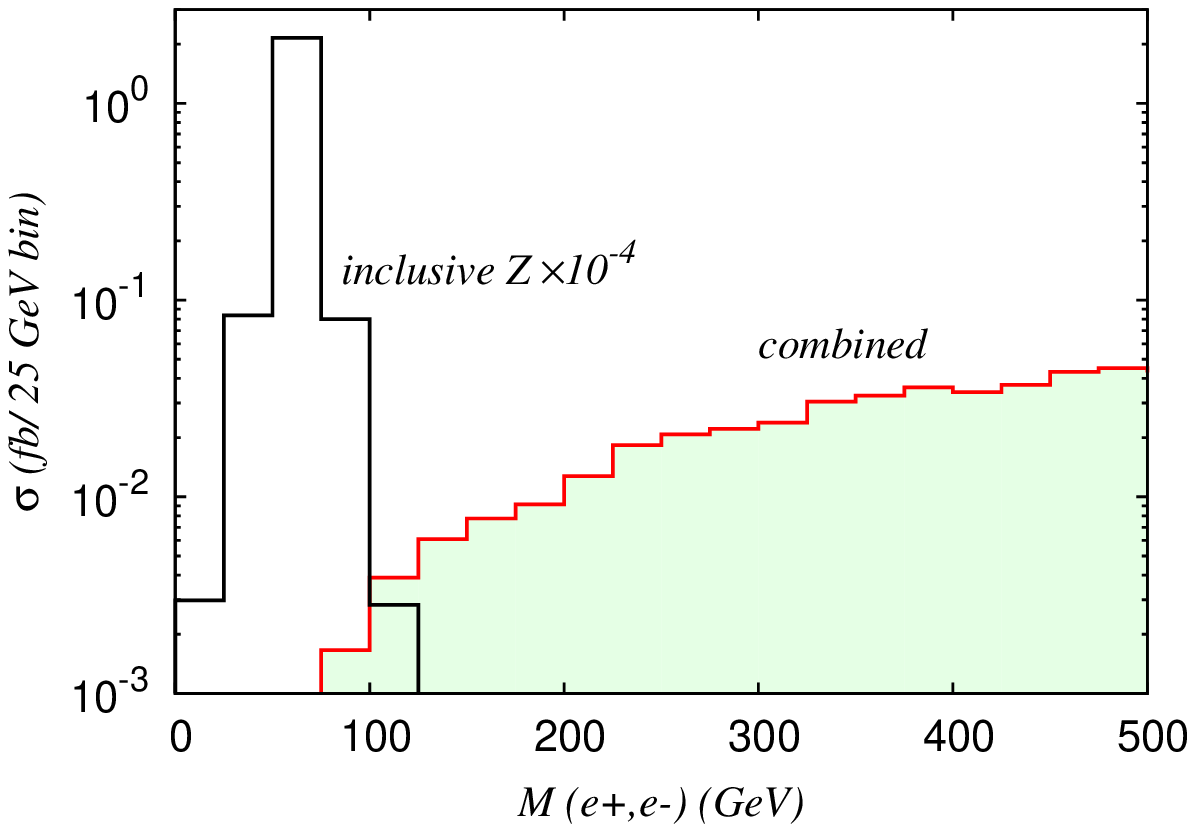}} &&
\resizebox{65mm}{!}{\includegraphics{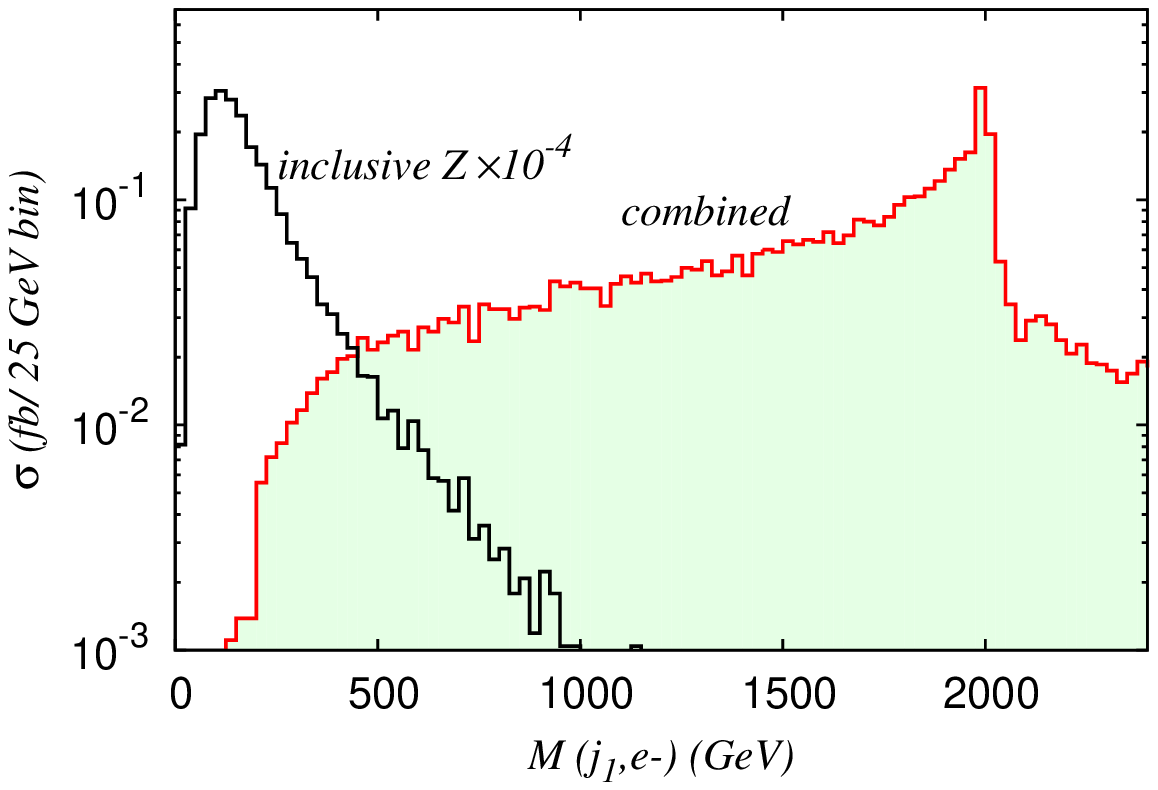}} \\
\footnotesize{\hspace{1.1cm}(c)}&&\footnotesize{\hspace{1.2cm}(d)}\\
\end{tabular}}
\caption{\label{fig:SigBGDist} Comparison between various distributions for 
the combined signal with $M_{e_8}=2$ TeV ($\Lm = 5$ TeV) and the inclusive 
$Z$ background for the 14 TeV LHC. The inclusive $Z$ background is shown after multiplying by 
a factor of $10^{-4}$. For each plot we indicate the corresponding bin size in the Y-axis label.}
\end{figure}

\begin{table}[!h]
\bc
\begin{tabular}{l c r }\hline
Process &~& Cross section (fb)  \\ 
\hline\hline  
$Z+nj$  && $2.11E4$ \\ 
 
$tt$    && $1.95E3$ \\ 

$tW$    && 132.15 \\ 

$WW$    && 7.51 \\
\hline 
Total   && $2.32E4$ \\
\hline
\end{tabular}
\caption{\label{tab:SMBG}The main SM backgrounds to the combined production of
$e_8$'s obtained after applying the Basic cuts (see text for definition).}
\ec
\end{table} 

\subsection{Kinematical Cuts}
In Fig. \ref{fig:SigBGDist}a we display the $p_T$ distributions of $e$'s from 
the combined signal and the inclusive $Z$ production, respectively. For the 
signal, we have chosen $M_{e_8} = 2$ TeV and $\Lm=5$ TeV. As expected, the 
distribution for the $e$ coming from the background has a peak at about $M_Z/2$ 
but there is no such peak for the signal. We can also see the difference between
the $p_T$ distributions of the leading $p_T$ jets for the signal and the 
background in Fig. \ref{fig:SigBGDist}b. We also display
the distributions of $M(e^+,e^-)$ (see Fig.\ref{fig:SigBGDist}c) and $M(e^-,j_1)$ 
(see Fig. \ref{fig:SigBGDist}d) (where $j_1$ denotes the leading $p_T$ jet) 
which show very different natures for the signal and the background.

Motivated by these distributions, we construct some kinematical cuts to separate 
the signal from the background:
\begin{enumerate}

\item {\bf Basic cuts}\\
For $x,y = e^+,e^-,j_1,j_2$ ($j_1$ and $j_2$ denote the first two of the $p_T$ ordered jets respectively),

\begin{enumerate}
\item
$p_T(x) > 25$ GeV
\item
Rapidity, $|\eta(x)| < 2.5$
\item
Radial distance, $\Delta R(x,y)_{x\neq y} \geq 0.4$
\end{enumerate}

\item {\bf Discovery cuts}
\begin{enumerate}
\item
All the \emph{Basic} cuts
\item
$p_T(e^{+}/e^{-}) > 150$ GeV; $p_T(j_1) > 100$ GeV
\item 
$M(e^+,e^-) > 150$ GeV
\item For at least one combination of $(e,j_i)$:
$|M(e,j_i) - M_{e_8}| \leq 0.2 M_{e_8}$ where $e =e^+$ or $e^-$ and $j_i=j_1$ or $j_2$.
\end{enumerate}
\end{enumerate}
The cut on $M(e^{+},e^{-})$ can remove the $Z$ inclusive
background almost completely. For our estimation of the LHC discovery reach we also use the cut on $M(e,j_i)$ to demand that either of the electrons 
reconstruct to an $e_8$ when combined with any one of $j_1$ or $j_2$.  Although this cut involves an unknown parameter, namely $M_{e_8}$, it can be implemented in the actual experiment by performing a scan of  $M_{e_8}$ over a range (say, 0.5 TeV to 4 TeV). While scanning, for each value of $M_{e_8}$, one can apply this cut on all the events. If $e_8$ exists within the scanned region, it will lead to an excess of events (compared to the SM) around the actual value of $M_{e_8}$.
We find that the Discovery cuts can reduce the SM background
drastically. Especially for higher $M_{e_8}$ the background becomes much smaller
compared to the signal, making it essentially background free\footnote{
For $M_{e_8}=0.5$ TeV (1 TeV) we estimate the total SM background 
with the Discovery cuts at the 14 TeV LHC to be about 4 fb (0.3 fb). Although
these numbers are only rough estimates for the actual SM backgrounds (as, 
{\it e.g.}, we don't consider the effect of any loop induced diagrams) they 
indicate the SM backgrounds become very small compared to the signal (see Table
\ref{signal}) after the Discovery cuts.}. In Table \ref{signal} we show the
signal with the above two cuts applied.

\begin{table}
\centering
\begin{tabular}{|c|r|r|r|r|}
\hline
\multicolumn{1}{|c|}{$M_{e_8}$} & \multicolumn{2}{c|}{$\Lm = 5$ TeV} & \multicolumn{2}{c|}{$\Lm = 10$ TeV}\\
\cline{2-5}
 (GeV) & Basic (fb) & Disco. (fb) & Basic (fb) & Disco. (fb)  \\ 
\hline\hline
500		& 2.73E4  & 1.31E4  & 2.70E4 & 1.27E4 \\

750		& 2.63E3  & 1.93E3  & 2.59E3 & 1.91E3 \\

1000    & 442.95  & 367.20  & 415.35 & 347.16 \\ 

1250	& 105.21  &  90.25  &  91.99 &  80.45 \\

1500    &  31.73  &  27.25  &  24.54 &  21.86 \\ 

1750	&  11.53  &   9.76  &   7.52 &   6.71 \\

2000    &   4.77  &   3.92  &   2.59 &   2.28 \\
 
2250	&   2.26  &   1.80  &   0.99 &   0.85 \\

2500    &   1.18  &   0.91  &   0.42 &   0.36 \\
 
2750 	&   0.65  &   0.49  &   0.20 &   0.16 \\

3000    &   0.37  &   0.27  &   0.11 &   0.08 \\ 

3250	&   0.22  &   0.16  &   0.06 &   0.04 \\

3500	&   0.13  &   0.09  &   0.03 &   0.02 \\
\hline 
\end{tabular} 
\caption{\label{signal}The combined signal after basic and Discovery cuts (see text 
for the definitions of the cuts) for $\Lm = 5$ TeV and 10 TeV for different $M_{e_8}$.}
\end{table}

\subsection{LHC Reach with Combined Signal}

\begin{figure}[!h]
\centering
\subfloat{
\begin{tabular}{c}
\resizebox{90mm}{!}{\includegraphics{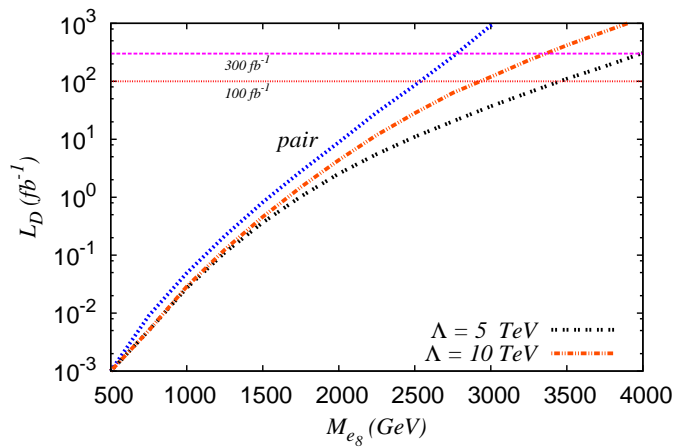}} \\
\end{tabular}}
\caption{\label{fig:Reach} The required luminosity for discovery ($L_D$) as a
function of $M_{e_8}$ with $\Lm = 5$ TeV and 10 TeV at the 14 TeV LHC for 
combined production with Discovery cuts (see text for the definitions of 
the cuts). The $L_D$ for pair production is computed after demanding two
$e_8$'s are reconstructed instead of one.}
\end{figure}

We define the luminosity requirement for the discovery of color octet electrons 
at the LHC as following:
\ba
L_{D} &=&  Max(L_5,L_{10})
\ea
where $L_5$ denotes the luminosity required to attain $5\s$ statistical
significance for $S/\sqrt{B}$ and $L_{10}$ is the luminosity required to
observe 10 signal events. We show $L_{D}$ as a function of $M_{e_8}$ for the
Discovery cuts in Fig. \ref{fig:Reach} for $\Lm=5$ TeV and 10 TeV at the 
14 TeV LHC. In Fig. \ref{fig:Reach} we also plot the $L_D$ using only the pair
production process. To estimate the pair production from the combined signal we 
apply a set of kinematical cuts almost identical to the Discovery cuts
except that now we demand that the two electrons and the two leading $p_T$ jets
reconstruct to two $e_8$'s instead of one:
\begin{enumerate}
\item {\bf Pair production extraction cuts}
\begin{enumerate}
\item
All the \emph{Basic} cuts
\item
$p_T(e^{+}/e^{-}) > 150$ GeV; $p_T(j_1) > 100$ GeV
\item 
$M(e^+,e^-) > 150$ GeV
\item
$|M(e^+,j_k) - M_{e_8}| \leq 0.2M_{e_8}$ and $|M(e^-,j_l) - M_{e_8}| \leq 0.2M_{e_8}$
with  $k\neq l= \{1,2\}$.
\end{enumerate}
\end{enumerate}

In Fig. \ref{fig:Reach}, $L_D$ goes as $L_{10}$ for both pair and combined
productions, as in these cases the backgrounds become quite small compared to 
the signals. With the Discovery cuts the reach goes up to 3.4 TeV and 2.9 TeV 
(4 TeV and 3.3 TeV) with 100 fb$^{-1}$ (300 fb$^{-1}$) integrated luminosity for
$\Lm=5$ TeV and 10 TeV respectively at the 14 TeV LHC. This also shows that for
$\Lm=5$ TeV (10 TeV) with combined signal at 14 TeV LHC with 300 fb$^{-1}$ 
integrated luminosity the reach goes up from the pair production by almost 1.2 
TeV (0.5 TeV). However, we should keep in mind that this increase depends on $\Lm$.
As the single production cross section goes like $1/\Lm^2$, if $\Lm$ is smaller 
than 5 TeV then the reach of the combined production will increase even more but 
for higher $\Lm$ (like $\Lm=10$ TeV as shown in Fig. \ref{fig:Reach}) its $L_D$ 
plot will approach more towards the pair production plot.

\section{Summary and Conclusions}
\label{sec:conclu}

In this paper we have studied the phenomenology of $e_8$'s at the LHC and 
estimated the discovery potential of  such particles at the LHC. We have 
explored various production channels of $e_8$'s at the LHC namely, the pair, 
the two-body single, the three-body single and the indirect production channels. 
The contribution  of the three-body single production channel 
is comparable to that of the two-body single production channel. While the pair production 
cross section dominates the other channels for low $M_{e_8}$, for high values 
of $M_{e_8}$ the single productions become significant. The 14 TeV LHC with 
100 fb$^{-1}$ (300 fb$^{-1}$) of integrated luminosity can probe $e_8$'s with 
masses up to 2.5 TeV (2.8 TeV) 
with only pair production. We have demonstrated how this reach can be increased
further by combining signal events from different production processes. However,
this increment is $\Lm$ dependent as the single production cross section scales 
as $1/\Lm^2$. For $\Lm=5$ TeV (10 TeV) the increment is about 0.9 TeV (0.4 TeV)
with 100 fb$^{-1}$ of integrated luminosity at the 14 TeV LHC and with 
300 fb$^{-1}$ of integrated luminosity it is about 1.2 TeV (0.5 TeV).

We point out that  our analysis can also be used to probe $\Lm$, the compositeness scale, for any fixed $M_{e_8}$. This is possible because of the scaling 
of the single production cross section with $\Lm$. For example, for $M_{e_8}=2$ TeV
and $\Lm=10 $ TeV we estimate the single production cross section as $1.2$ fb 
which we obtain from the events that pass the Discovery cuts but not the Pair
production extraction cuts. By computing $L_{10}$ ({\it i.e.}, the luminosity
requirement to observe 10 signal events) from this we can conclude that for
$M_{e_8}=2$ TeV the 14 TeV LHC with 100 fb$^{-1}$ (300 fb$^{-1}$) of integrated
luminosity can probe $\Lm \sim 35$ TeV ($55$ TeV). 
This can be useful as the present searches of  leptoquarks at the LHC generally focus on their pair production which is mostly independent of $\Lm$. However, even with the single production of leptoquarks at the LHC it is difficult to probe the compositeness scale directly. This can be seen in an effective theory picture as, unlike leptogluon, the interaction Lagrangian of leptoquarks with the SM particles is 
dominated by renormalizable dimension four operators.

We note that the data from the current leptoquark searches at the LHC
can be used to search for $e_8$'s also. The current data for leptoquarks
\cite{Chatrchyan:2012dn,Chatrchyan:2012sv} already puts some constraints 
on the masses of the $l_8$'s. For example the data from the search  for 1$^{st}$
generation charged leptoquark in the pair production channel clearly rules out a
color octet electron of mass less than 900 GeV (see Fig. 10 of
\cite{Chatrchyan:2012dn}), since the pair production cross section of color octet
electrons is always larger than the pair production cross section of color triplet
leptoquarks of the same mass due to color enhancement \footnote{After our paper was
posted in the arXiv, a paper appeared \cite{Goncalves-Netto:2013nla}  where the
authors estimate the exclusion limit of charged leptogluons from the CMS leptoquark
data to be 1.2-1.3 TeV.}.

\section*{Acknowledgements}
We thank J. Alwall and F. Maltoni for helping us with matching. S. M. thanks 
R. Barcelo for helpful comments.

\appendix

\section{Preparation of Matched Signal}\label{matching}

For the signal, we match the matrix element partons with the parton showers 
using the shower-$k_T$ scheme \cite{Alwall:2008qv}  in MadGraph5 with the 
matching scale $Q_{cut}\sim 50$ GeV. We generate the combined signal including 
the different production processes as discussed in section \ref{sec:LHCdis},
\ba
pp &\xrightarrow{e_8} & ee \textrm{ (includes $P_{ind} $)}\nonumber\\
pp &\xrightarrow{e_8} & ee + 1\textrm{-j (includes $P_{ind} +$ 1-j, $P_{2Bs}$ )}\nonumber\\
pp &\xrightarrow{e_8} & ee + 2\textrm{-j (includes $P_{ind} +$ 2-j, $P_{2Bs}+$ 1-j, $P_{pair}$ , $P_{3Bs}^3$ )}\nonumber\\
pp &\xrightarrow{e_8} & ee + 3\textrm{-j (includes $P_{ind} +$ 3-j, $P_{2Bs}+$ 2-j, $P_{pair}+$ 1-j, $P_{3Bs}^3+$ 1-j)}
\ea
where $P_{pair}$, $P_{2Bs}$, $P_{3Bs}^3$ and $P_{ind}$ are the pair, two-body
single, three-body single of the third type and indirect productions respectively.
We refer the reader to \cite{Alwall:2008qv} and the references therein for more
details on the matching scheme and the procedure.

\end{document}